\documentclass[aps,prd,preprint,showpacs,groupedaddress,nofootinbib]{revtex4}
\usepackage{amssymb,amsbsy,color}
\usepackage{latexsym}
\usepackage{graphics}
\usepackage{graphicx}
\usepackage{amsmath}

\newcommand{\be}{\begin{equation}}
\newcommand{\ee}{\end{equation}}
\newcommand{\bea}{\begin{eqnarray}}
\newcommand{\eea}{\end{eqnarray}}
\newcommand{\ba}{\begin{array}}
\newcommand{\ea}{\end{array}}
\newcommand{\bal}{\begin{align}}
\newcommand{\eali}{\end{align}}
\newcommand{\nn}{\nonumber}

\newcommand{\pref}[1]{(\ref{#1})}

\newcommand{\lt}{\overline{l}_3}
\newcommand{\lf}{\overline{l}_4}

\begin{document}
\preprint{HU-EP-10/44, SFB-CCP-10-72}
\title{
Chiral logs in twisted mass lattice QCD with large isospin breaking
}
\author{$^{1}$Oliver B\"ar}

\affiliation{
$^1$Institute of Physics, Humboldt University Berlin, Newtonstrasse
15, 12489 Berlin, Germany
}

\date{\today}

\begin{abstract}

The pion masses and the pion decay constant are calculated to 1-loop order in SU(2) twisted mass Wilson chiral perturbation theory, assuming a large pion mass splitting and tuning to maximal twist. Taking the large mass splitting at leading order in the chiral expansion leads to significant modifications in the chiral logarithms. For example, the result for the charged pion mass contains a chiral logarithm that involves the neutral pion mass instead of the charged one. Similar modifications appear in the results for the neutral pion mass and the decay constant. 
These new results are used in fits to lattice data obtained recently by the European twisted mass collaboration. The data can be fitted well, in general better than with the standard chiral perturbation theory expressions that ignore the mass splitting. The impact on the extraction of low-energy couplings is briefly discussed.

\end{abstract}
\maketitle

\section{Introduction}

Lattice QCD with twisted mass Wilson fermions \cite{Frezzotti:2000nk,Frezzotti:2001ea} has some advantages compared
to its counterpart with a standard mass term. The most prominent one certainly is automatic O($a$) improvement at maximal twist \cite{Frezzotti:2003ni}. An efficient algorithm \cite{Urbach:2005ji} allows simulations with sufficiently small pion masses to be in the chiral regime of QCD. Many results have been obtained in the quenched approximation as well as for unquenched 2-flavor-QCD (for a review see ref.\ \cite{Shindler:2007vp}). Recently, first results with additional active strange and charm quarks (2+1+1 flavor simulations \cite{Frezzotti:2004wz}) have been reported too \cite{Baron:2010bv}. 

A drawback of the twisted mass formulation is the breaking of isospin symmetry, most clearly seen in a mass splitting between the charged and neutral pions. This breaking is not a fundamental concern; it is a lattice artifact of O($a^2$)  and isospin symmetry is automatically restored in the continuum limit. Nevertheless, at nonzero lattice spacing the mass splitting is rather large. For example, for the 2+1+1 flavor simulations at $a\approx 0.078$fm one finds $M_{\pi^0}/M_{\pi^{\pm}}\approx 0.54$ for $M_{\pi^{\pm}}\approx 320$MeV and $M_{\pi^0}/M_{\pi^{\pm}}\approx 0.77$ for $M_{\pi^{\pm}}\approx 400$MeV.\footnote{Numbers from tables 5 and 8 of ref.\ \cite{Baron:2010bv}.} The neutral pion masses have a ten percent error; still, the neutral pion mass is roughly half as heavy as the charged one for the smaller charged pion mass. This is not a small effect.

A large splitting in the pion masses is worrisome because of the chiral extrapolation which needs  to be performed in order to get results at the physical pion mass. 
The standard tool here is chiral perturbation theory (ChPT) \cite{Weinberg:1978kz,Gasser:1983yg,Gasser:1984gg}. This low-energy effective field theory of QCD provides the quark mass dependence of observables. In particular, it predicts a characteristic non-analytic quark mass dependence, so-called chiral logarithms (chiral logs for short). In continuum ChPT without isospin breaking the chiral logs involve degenerate charged and neutral pion masses. The same is true for the Wilson ChPT (WChPT) expressions in ref.\ \cite{Sharpe:2004ny}, which incorporate the lattice spacing corrections assuming to be in the GSM regime where $\mu\sim a\Lambda_{\rm QCD}^2$. In fact, at maximal twist the NLO expressions in the GSM regime are identical to those in the continuum.
However, if the mass splitting is large, one can expect results involving the logs $M^2_{\pi^{\pm}}\ln M^2_{\pi^{\pm}}/\Lambda^2$ and $M^2_{\pi^{0}}\ln M^2_{\pi^{0}}/\Lambda^2$. This modification may have a non-negligible impact on the chiral extrapolation. The extraction of the Gasser-Leutwyler (GL) coefficients, which are associated with the chiral logs, may be strongly affected by a large mass  splitting. It is even possible that the modifications are so severe that the lattice data are not described at all by the continuum ChPT results.
 
All this is reminiscent of staggered fermions with taste symmetry breaking. The lattice simulations by the MILC collaboration with Asqtad staggered fermions show a sizeable mass splitting between the various taste partners.\footnote{For a comprehensive review see ref. \cite{Bazavov:2009bb}.} For example, the ratio of the heaviest (taste singlet) pion mass $M_{\pi_I}$ and the lightest (Goldstone) pion mass $M_{\pi_5}$ is approximately 0.77 at $a\approx 0.09$fm for a Goldstone pion mass $ M_{\pi_5} \approx 320$MeV \cite{Aubin:2004wf}.\footnote{Numbers from tables III and IV of ref.\ \cite{Aubin:2004wf}.} These large taste splittings have been included in the chiral logarithms \cite{Bernard:2001yj,Aubin:2003mg,Aubin:2003uc} (so-called staggered ChPT), and the lattice data clearly favor these modified logs. In fact, the continuum ChPT expressions cannot be fitted at all to the lattice data.
     
In this paper we compute the pion masses (both for the charged and the neutral pion) and the decay constant to 1-loop order in SU(2) WChPT with the mass splitting  taken into account in the chiral logs. In the language of WChPT we work in the Aoki regime \cite{Aoki:2003yv} (also called LCE regime). For simplicity we work at maximal twist only, which is the relevant case in practice. As expected, we do find deviations from the continuum ChPT results. For example, the 1-loop result for the charged pion mass has a chiral log proportional to $M^2_{\pi^0}\ln  M^2_{\pi^0}/\Lambda^2$, i.e.\ it involves the neutral pion mass. The continuum chiral log proportional to $M_{\pi^{\pm}}^2\ln  M^2_{\pi^{\pm}}/\Lambda^2$, on the other hand, is not present at all.\footnote{Also, this finding has its analogue in staggered ChPT. The 1-loop result for the Goldstone pion mass has chiral logs involving the taste partners $M_{\pi_I}, M_{\pi_V}, M_{\pi_A}$. The naively expected chiral log involving $M_{\pi_5}$ is absent \cite{Aubin:2003mg}.} 
Similar changes are found for the neutral pion mass and the pion decay constant.

There is a second aspect to these modifications that is equally important. 
A smaller neutral pion mass implies larger finite volume (FV) corrections. 
Typically, the FV corrections are exponentially suppressed with $M_{\pi}L$, where $L$ is the spatial extent of the finite volume \cite{Gasser:1986vb,Gasser:1987ah,Gasser:1987zq}. If the neutral pion is significantly lighter than the charged one, the associated FV corrections are significantly less suppressed. Formulated the other way around one can say that the FV corrections due to the neutral pion mass are exponentially enhanced. This has been pointed out recently in ref.\ \cite{Colangelo:2010cu}, where the enhanced FV corrections have been computed using the resummed L\"uscher formula \cite{Colangelo:2004xr,Colangelo:2005gd,Colangelo:2006mp}. This enhancement can be large. For example, consider again the aforementioned lattice data with $M_{\pi^0}/M_{\pi^{\pm}}\approx 0.54$. The volume is such that $M_{\pi^{\pm}}L \approx 4.0$, which implies $M_{\pi^{0}}L \approx 2.2$ and significantly larger FV corrections.

In the last section of this paper we show that the most recent  lattice data of the ETM collaboration are indeed sensitive to the modifications in the chiral logs. We reanalyze the data in ref.\ \cite{Baron:2010bv} using our new results. Indeed, the data prefer the results with a neutral pion mass smaller than the charged one, and the pion mass splitting is compatible with the one directly measured. More importantly, the central values for some of the extracted LECs change sizably with our new fit formulae. 

The rest of this paper is organized as follows. In section II we briefly summarize various results of twisted mass WChPT that we need later on. The primary purpose is to settle our notation and to prepare the 1-loop calculation presented in section III. The following section IV contains the analysis of the 2+1+1 flavor data in ref.\ \cite{Baron:2010bv} using our newly derived results. Final conclusions are drawn in section V.

\section{Twisted mass Wilson ChPT}

Wilson ChPT (WChPT) \cite{Sharpe:1998xm,Rupak:2002sm} is the low-energy effective theory for lattice QCD with Wilson quarks. It is based on a joint expansion in small pion momenta, quark masses and small lattice spacings $a$. Pedagogical introductions to WChPT are given in refs.\ \cite{SharpeNara,GoltLesHouches}, where many references to the original literature can be found as well.

WChPT has two sources of explicit chiral symmetry breaking, the quark mass and the lattice spacing, and the appropriate power counting depends on their relative size. The literature distinguishes two different regimes which seem to be the relevant ones for present-day lattice simulations. The GSM ({\em generically small masses}) regime \cite{Sharpe:2004ny} assumes that the breaking of chiral symmetry due to the quark mass and the lattice spacing is of equal size, $m\sim a \Lambda^2_{\rm QCD}$. The Aoki or LCE ({\em large cut-off effects}) regime \cite{Aoki:2003yv,Aoki:2008zk}, on the other hand, assumes smaller quark masses or larger lattice spacings such that $m\sim a^2 \Lambda^3_{\rm QCD}$. This is the regime we focus on in this paper, because it implies a pion mass splitting of the order of the charged pion mass itself.
In the following we collect a few results that have been published before in various papers \cite{Aoki:2004ta,Sharpe:2004ny,Aoki:2006nv}; the main purpose is to introduce our notation.

The leading order (LO) chiral lagrangian in the LCE regime reads
\begin{align}
{\cal L}_{\rm LO} &= {\cal L}_2 + {\cal L}_{a^2}\,.
\end{align}
Here ${\cal L}_2$ denotes the standard LO lagrangian from continuum ChPT
\cite{Gasser:1983yg,Gasser:1984gg},
\begin{align}\label{L2m}
{\cal L}_2 &= \frac{f^2}{4} \langle \partial_{\mu}\Sigma \partial_{\mu}\Sigma^{\dagger} \rangle - \frac{f^2B}{2}\langle \Sigma M^{\dagger} + M \Sigma^{\dagger} \rangle \,.
\end{align}
$\langle\ldots\rangle$ stands for the trace over the flavor indices.
$\Sigma$ denotes the usual SU(2)-valued Goldstone boson field which involves the pion fields in the standard way, specified explicitly below. $f$ and $B$ are the familiar LO low-energy coefficients (LECs).\footnote{With our conventions the decay constant $f_{\pi}\approx f \approx$ 93MeV.} $M$ denotes the quark mass matrix which in the presence of a twisted mass $\mu$ has the form
\begin{align}
M&= m + i\mu\sigma_3\,,\label{masstermtwistedchiral} 
\end{align}
where $\sigma_3$ is the third Pauli matrix.
Note that the standard (untwisted) mass $m$ refers to the so-called shifted mass  which includes the O($a$) shift to the additive mass renormalization \cite{Sharpe:1998xm}. 
The lagrangian ${\cal L}_{a^2}$ contains the leading O$(a^2)$ correction \cite{Bar:2003mh} and reads  (we follow the notation of ref.\ \cite{Sharpe:2004ny})
\begin{align}
{\cal L}_{a^2}  &= -\hat{a}^2 W^{\prime}_{68} \langle  \Sigma^\dag + \Sigma \rangle^2 
 \,.\label{laga2SU2}
\end{align}
$W^{\prime}_{68}$ is the LEC associated with the O$(a^2)$ correction and $\hat{a}$ is defined by $ \hat{a}=2W_0 a$.
$W_0$ is a LEC of mass dimension three \cite{Rupak:2002sm}, hence $\hat{a}$ has dimension two and $W^{\prime}_{68}$ is dimensionless. 

The SU(2) flavor symmetry is explicitly broken by the twisted mass term `pointing' into the $\sigma_3$ direction. Consequently, the ground state $\Sigma_{\rm vac}$, i.e.\ the minimum of the classical potential energy, is no longer equal to the identity but of the form \cite{Munster:2004am,Scorzato:2004da,Sharpe:2004ps}
\begin{align}\label{Ansatzvacangle}
\Sigma_{\rm vac} &= \exp(i\phi \sigma_3)\,,
\end{align}
where $\phi$ is called the vacuum angle. It is determined by minimizing the potential energy which reads
\begin{align}\label{potenergySU2twist}
V &= -\frac{f^2B}{2}\langle M^{\dagger}\Sigma + \Sigma^{\dagger}M \rangle + \frac{f^2}{16}c_2a^2 
\langle \Sigma + \Sigma^{\dagger} \rangle^2\,.
\end{align} 
For notational convenience we introduced the short hand notation\footnote{Note that our definition for $c_{2}$ is not exactly the same as in  \cite{Sharpe:1998xm}.  It  differs by a factor of $f^{2}a^{2}$ and we have dropped the terms proportional to the quark mass.}
\begin{align}
\label{c2}
c_2 & =  - 64W_{68}^{\prime}\frac{W_0^2}{f^2}\,.
\end{align}
This combination of LECs is of mass dimension four.
Using the ansatz \pref{Ansatzvacangle} in $V$ we can derive the gap equation ${\rm d}V/{\rm d}\phi=0$, which can be written as \cite{Aoki:2004ta}
\begin{align}\label{GapEq_final}
2B \mu\cos\phi &=\sin\phi\left(2Bm-2c_2 a^{2}\cos\phi\right) .
\end{align}
This equation determines the vacuum angle as a function of the variable parameters in the theory, the two masses $m,\mu$ and the lattice spacing $a$: $\phi=\phi(m,\mu,a)$. 
Solutions of the gap equation and the associated phase diagram of the theory are discussed at length in the literature \cite{Sharpe:1998xm,Munster:2004am,Sharpe:2004ps}. It turns out that there are two qualitatively different scenarios depending on the sign of $c_2$. 
For $c_2>0$ there exists an Aoki phase \cite{Aoki:1983qi} for $\mu=0$. Parity and flavor are spontaneously broken in this phase, and the charged pions are massless due to the spontaneous breaking of the flavor symmetry. Negative values of $c_2$, on the other hand, imply a first order phase transition at $m=0$. The pions are always massive except for the neutral pion which becomes massless at the end of the phase transition line, given by $|2B\mu |= -2c_2a^2$. 

As usual, the tree level pion masses are calculated by expanding the field $\Sigma$ around the vacuum configuration. We parametrize $\Sigma$ according to 
\begin{align}
\Sigma(x) = \Sigma_{\rm vac}^{1/2} \exp\left(\sum_{k=1}^3 i\pi_k(x)\sigma_k/f\right)\Sigma_{\rm vac}^{1/2}\,.
\end{align}
Using this form in \pref{potenergySU2twist} and making use of the gap equation the LO pion masses are found as \cite{Aoki:2004ta}
\begin{align}
M^2_{\pi^\pm}&=\frac{2B\mu}{\sqrt{1-t^2}}\,,\label{charMpi}\\
M^2_{\pi^0} &=\, M^2_{\pi^\pm} + \Delta M_{\pi}^2\,,\quad\quad \Delta M_{\pi}^2\,=\,
2c_2a^2(1-t^2)\,, \label{neutMpi}
\end{align} 
where we introduced
\begin{align}
t&= \cos \phi\,.
\end{align} 
However, for some special values of $t$ one has to be careful. Obviously, the charged pion mass seems ill-defined for $t=\pm 1$, but the gap equation immediately tells us that $t=\pm 1$ is a solution only if $\mu=0$. This is the standard untwisted case and one finds (for positive $c_2$ and outside the Aoki phase) $M^2_{\pi^\pm}=2Bm + 2c_2a^2 = M^2_{\pi^0}$ \cite{Aoki:2004ta}.  Another case that requires care is $t=0$ and $c_2<0$. It seems that the squared neutral pion mass can become negative for sufficiently small $\mu$ values. However, in this case the result \pref{neutMpi} is only valid for $|2B\mu|\ge-2c_2a^2$, keeping $M^2_{\pi^0}$ always non-negative. As mentioned before, $|2B\mu |= -2c_2a^2$ corresponds to the endpoints of the phase transition line with vanishing neutral pion mass.

So far the results in \pref{charMpi} and \pref{neutMpi} are valid for arbitrary $m,\mu$ and $a$ (recall the implicit dependence on these parameters via $t(m,\mu,a)$). Usually one is interested in the special case of maximal twist only. Here maximal twist refers to the situation where the untwisted mass assumes a particular (`critical') value, $m=m_{\rm cr}$. The most popular and widely used definition specifies maximal twist as the point where the PCAC quark mass $m_{\rm PCAC}$ vanishes (so-called PCAC mass definition). 
The PCAC quark mass is defined by 
\begin{align}
m_{\rm PCAC} &= \frac{\langle \partial_{\mu}A_{\mu}^c(x) P^c(y)\rangle}{ 2 \langle P^c(x) P^c(y)\rangle}
\end{align}
with flavor index $c=1,2$. The expressions for the axial vector current $A_{\mu}^c$ and the pseudo scalar density $P^c$ have been worked out in refs.\ \cite{Sharpe:2004ny,Aoki:2009ri}. To LO these are the familiar expressions from continuum ChPT plus corrections proportional to powers of the lattice spacing.\footnote{Note that the currents and densities can be given either in the twisted or the physical basis, which are related by a field transformation. Here we always work in the twisted basis.} As any observable, the PCAC quark mass is a function of $m,\mu$ and $a$. Hence, cast into an equation the PCAC mass definition for maximal twist reads, reads
\begin{align}
m_{\rm PCAC}(m=m_{\rm cr}^{},\mu,a) &= 0.\label{pcacmasscondition}
\end{align}
Note that $m_{\rm cr}$ does in general depend on the twisted mass and the lattice spacing: $m_{\rm cr}^{}(\mu,a)$. 

The PCAC mass is easily computed at tree level \cite{Aoki:2004ta,Sharpe:2004ny,Aoki:2006nv}. To LO in the LCE regime one finds \cite{Aoki:2004ta}
\begin{align}
m_{\rm PCAC} &= m - \frac{c_2a^2}{B} t\,.
\end{align}
This vanishes for $m=0$ (which implies $t=0$). For this particular value the result \pref{charMpi} for the charged pion mass turns into the one from continuum ChPT, with $m$ replaced by $\mu$. The pion mass splitting between the neutral and charged pion mass is maximal in this case.

Taking into account higher order terms in the calculation of the PCAC mass \cite{Sharpe:2004ny,Sharpe:2005rq,Aoki:2006nv} one finds that a vanishing PCAC mass implies $t={\rm O}(a)$. Dropping all terms higher than linear in $a$ we can approximately write $t=X a$, where $X$ is some combination of LECs. Note that this result still guarantees automatic O($a$) improvement. Using $t=X a$ in \pref{charMpi}, \pref{neutMpi} and expanding the square root we find small O($a^2$) corrections to the result for $t=0$. Moreover, the same is true for the terms stemming from the O($ap^2,a\mu)$ terms in the chiral lagrangian \cite{Sharpe:2004ny,Aoki:2006nv}. All these terms are associated with one power of $t$. Hence, at maximal twist with $t={\rm O}(a)$ these terms become scaling violations of O($a^2$), in agreement with automatic O($a$) improvement at maximal twist.

In the following we always assume to be in  the LCE regime.
The precise meaning of this assumption is that the contributions $M^2_{\pi^{\pm}}$ and $\Delta M_{\pi}^2$ to the neutral pion mass in \pref{neutMpi} are of the same order, $M^2_{\pi^{\pm}} \sim \Delta M_{\pi}^2$. At maximal twist this is equivalent to $\mu\sim c_2a^2/B$. Assuming that all dimensionful constants are roughly equal to some power of $\Lambda_{\rm QCD}$ we recover the condition we have given before: $\mu\sim a^2\Lambda_{\rm QCD}^3$.

\section{Pion masses and decay constant to one loop}

\subsection{Propagators, vertices and counterterms}

Our goal is to compute the 1-loop corrections to the tree level pion masses given in \pref{charMpi} and \pref{neutMpi}. In order to simplify the calculation we restrict ourselves to maximal twist, which is the relevant case in practice. We keep the ${\cal L}_{a^2}$ lagrangian at LO, hence the 1-loop calculation takes into account  the following terms:
\begin{align}
\begin{array}{rcl}
{\rm LO:}& \quad & p^2,\,M,\,a^2\\
{\rm NLO:}& \quad & p^4,\,p^2M,\,M^2,\,p^2a^2,\, M a^2,\,a^4
\end{array}
\end{align}
Here the restriction to maximal twist implies already some simplification: In general the chiral Lagrangian contains also terms of O($p^2a,\,M a)$ \cite{Sharpe:2004ny} as well as O($a^3$). However, these are proportional to the solution $t$ of the gap equation. Hence, at maximal twist with $t={\rm O}(a)$ these terms are promoted to O($p^2a^2, M a^2,a^4$) terms, and the list given above is meant to include these NLO corrections.

For the 1-loop calculation we need the propagators and interaction vertices stemming from the LO lagrangian. The propagators involve the tree level pion masses given in \pref{charMpi} and \pref{neutMpi}. In the 1-loop correction we can set $t=0$, since $t={\rm O}(a)$ leads to O$(a^2)$ corrections (to the chiral logs) which are beyond NLO. Therefore, the (Euclidean space-time) propagators in momentum space read
\begin{align}
G^{ab}(p^2) &= \frac{\delta^{ab}}{p^2 +M_{\pm}^2}\,,\qquad M_{\pm}^2\,=\,2B\mu\,,\qquad a,b=1,2\,,\label{Propchar}\\
G^{33}(p^2) &= \frac{1}{p^2 +M_{0}^2}\,,\qquad M_{0}^2\,=\,2B\mu +2c_2a^2\,.\label{Propneut}
\end{align}
The interaction vertices are the terms involving more than two pion fields. The four-pion vertices from the kinetic part in ${\cal L}_2$ are the same as in untwisted continuum ChPT. Expanding to quartic order in the pion fields we find
\begin{align}\label{VertexLkin}
{\cal L}_{p^2,4\pi}&=\frac{1}{6f^2}\Big((\partial_{\mu}\pi\cdot\pi)^2 - (\partial_{\mu}\pi)^2\pi^2\Big)\,.
\end{align}
Here we use the short hand notation $\partial_{\mu}\pi\cdot\pi=\sum_{c} \partial_{\mu}\pi_c\pi_c$, $(\partial_{\mu}\pi)^2=\sum_{c} \partial_{\mu}\pi_c\partial_{\mu}\pi_c$ and analogously for $\pi^2$.\footnote{We still keep the summation over the Lorentz index implicit.} 
Similarly, expanding the mass term and the lagrangian ${\cal L}_{a^2}$  to quartic order we find ($\pi^4 = (\pi^2)^2$)
\begin{align}
{\mathcal L}_{M,4\pi} & = -\frac{1}{24f^2} 2B(m \cos\phi +\mu\sin\phi)\pi^4\,,\label{Vertmassterm}\\
 {\mathcal L}_{c_2a^2,4\pi}& = \frac{1}{6f^2} 2c_2a^2\Big(\cos^2\phi \, \pi^4 -\sin^2\phi \,\pi^2\pi^2_3\Big)\,.\label{Verta2term}
\end{align}
It is more convenient to express the quark masses in \pref{Vertmassterm} by the tree level charged pion mass $M_{\pm}^2$ and the LO pion mass splitting $2c_2a^2$. With the help of the gap equation we can rewrite the sum of \pref{Vertmassterm} and \pref{Verta2term} as 
\begin{align}
{\mathcal{L}}_{M,4\pi}+{\mathcal{L}}_{c_2a^2,4\pi} & =  -\frac{1}{24f^2} M_{\pm}^2 \,\pi^4 + \frac{1}{8f^2} 2c_2a^2 t^2\,\pi^4
 -\frac{1}{6f^2} 2c_2a^2(1-t^2) \,\pi^2\pi^2_3\,.\label{Verttotal}
\end{align}
The vertices proportional to $a^2t^2$ lead, after contracting two of the four pion fields, to corrections proportional to $a^2t^2M^2_{\pi}\ln M^2_{\pi}/\Lambda^2$. For maximal twist with $t={\rm O}(a)$ these are corrections higher than NLO, hence they are ignored in the following. We can also ignore the three-pion vertices stemming from the mass term,
\begin{align}
{\mathcal L}_{M,3\pi} & = \frac{B}{3f} (\mu\cos\phi - m \sin \phi) \pi^2\pi_3\,.
\end{align}
Making again use of the gap equation this can be brought into 
\begin{align}
{\mathcal L}_{M,3\pi} & = - \frac{c_2a^2}{3f} t\sqrt{1-t^2} \pi^2\pi_3\,.
\end{align}
In order to form a loop diagram contributing to the self energy of the pions one needs two of these vertices. Therefore, with $t={\rm O}(a)$ this correction is at least proportional to $(c_2 a^2)^2 a^2 \sim a^6$, which is a NNLO correction in the LCE regime and beyond the order considered here. 

We conclude that  for an NLO calculation at maximal twist we can set $t=0$ in \pref{Verttotal}, and there is only one extra vertex proportional to $2c_2a^2$ besides the familiar ones from continuum ChPT. 

Finally, we need the analytic NLO corrections which also provide the necessary counter\-terms for the divergent contributions of the loop corrections. The full NLO lagrangian in the LCE regime at maximal twist consists of the following parts:
\begin{align}
{\cal L}_{\rm NLO} &= {\cal L}_4 + {\cal L}_{p^2a^2}+ {\cal L}_{Ma^2}+ {\cal L}_{a^4}+{\cal L}_{p^2a}+{\cal L}_{Ma}+{\cal L}_{a^3}
\,.
\end{align}
The first part ${\cal L}_4$ denotes the standard NLO lagrangian of continuum ChPT \cite{Gasser:1983yg} (we again follow the notation of \cite{Sharpe:2004ny}), 
\begin{align}
{\cal L}_4  &= {L_{45} \langle {\partial_{\mu} \Sigma\partial_{\mu} \Sigma^\dag  }
 \rangle \langle 
{\hat{M}  \Sigma^\dag + \Sigma \hat{M}^{\dagger} } \rangle }
- {L_{68} \langle {\hat{M}  \Sigma^\dag + \Sigma  \hat{M}^{\dagger} }
  \rangle ^2 }\,,\label{NLOct}
\end{align}
with 
\begin{align}
\hat{M}&=2BM
\end{align}
and the mass matrix $M$ defined in \pref{masstermtwistedchiral}. Note that we dropped all terms in ${\cal L}_4$ that do not contribute to the pion masses (and decay constant), in particular the O($p^4$) terms involving the GL coefficients $L_1,L_2$.

The next three parts in ${\cal L}_{\rm NLO}$ are constructed in appendix A of \cite{Aoki:2008zk}. Although a degenerate untwisted quark mass matrix was assumed in this reference  the generalization to the case with a twisted mass is straightforward. We find
\begin{align}
{\cal L}_{p^2a^2} &=  a_1 a^2 \langle \partial_{\mu}\Sigma  \partial_{\mu}\Sigma^{\dagger} \rangle + a_2 a^2 \langle \partial_{\mu}\Sigma  \partial_{\mu}\Sigma^{\dagger} \rangle \langle \Sigma+\Sigma^{\dagger}\rangle^2\nonumber\label{NLOp2a2}\\
&\phantom{=}\,\, + a_3 a^2 \langle \partial_{\mu}(\Sigma +\Sigma^{\dagger}) \rangle\langle \partial_{\mu}(\Sigma +\Sigma^{\dagger}) \rangle\,,\\[0.2ex]
{\cal L}_{Ma^2} &=  b_1 a^2 \langle \hat{M}^{\dagger} \Sigma + \Sigma^{\dagger}\hat{M}\rangle + b_2 a^2 \langle \Sigma + \Sigma^{\dagger}\rangle^2\langle \hat{M}^{\dagger} \Sigma + \Sigma^{\dagger}\hat{M}\rangle \,,\label{NLOMa2}\\[0.2ex]
{\cal L}_{a^4} &=  e_1 a^4 \langle \Sigma + \Sigma^{\dagger}\rangle^2 + e_2 a^4 \langle \Sigma + \Sigma^{\dagger}\rangle^4\,.\label{NLOa4}
\end{align}
The coefficients $a_j,b_j,e_j$ are undetermined LECs. 

The remaining corrections stem from the lagrangians 
\begin{align}
{\cal L}_{p^2a}&= W_{45}\hat{a}\langle \Sigma + \Sigma^{\dagger}\rangle\langle \partial_{\mu}\Sigma  \partial_{\mu}\Sigma^{\dagger} \rangle\,,\\
{\cal L}_{Ma}&= W_{68}\hat{a}\langle \Sigma + \Sigma^{\dagger}\rangle\langle \hat{M}^{\dagger} \Sigma + \Sigma^{\dagger}\hat{M}\rangle\,,\label{NLOeffMa}
\end{align}
given in ref.\ \cite{Sharpe:2004ny}, and the O($a^3$) correction 
\begin{align}\label{NLOa3}
{\cal L}_{a^3}&=d_1 a^3 \langle \Sigma + \Sigma^{\dagger}\rangle + d_2 a^3 \langle \Sigma + \Sigma^{\dagger}\rangle^3\,
\end{align}
given in ref.\ \cite{Aoki:2008zk}.\footnote{It is a matter of taste whether one uses $a$ or $\hat{a}$ in these expressions, the difference is just a multiplicative constant $2W_0$. However, the mass dimension of the LECs depends on this choice, because $a$ and $\hat{a}$  have mass dimension $-1$ and $2$, respectively.} 
Although in general of lower order in the chiral expansion than the ones discussed so far, these corrections are promoted to NLO terms at maximal twist. Let us demonstrate this for the contribution to  the pion masses. Expanding $\langle \Sigma + \Sigma^{\dagger}\rangle$ into pion fields we obtain
\begin{align}\label{MEleadingtot}
\langle \Sigma + \Sigma^{\dagger}\rangle &= t\Big(4 + \frac{2}{f^2}\pi^2\Big) +\ldots\,,
\end{align}
where the ellipses stand for the terms with three or more pion fields.\footnote{We dropped a term linear in $\pi_3$ as well, which does not play a role here.} Important is the factor $t$, which is of O($a$) at maximal twist. Setting $t=Xa$ with some constant $X$ we find a tree level correction of ${\cal L}_{Ma}$ to the pion masses proportional to $W_{68}Xa^2M^2_{\pm}$. This is a NLO correction in our counting and should be taken into account. Similar arguments can be made for the lagrangians ${\cal L}_{p^2a}$ and ${\cal L}_{a^3}$.

It is straightforward to expand all terms in \pref{NLOct} - \pref{NLOa3} to quadratic order in the pion fields. This leads to the analytic NLO corrections to the pion masses and wave function renormalization, which are of the generic form $A\pi^2/2$, $B\pi_3^2/2$ and $C(\partial_{\mu}\pi)^2/2$, with $A,B,C$ being some combinations of the LECs appearing in \pref{NLOct} - \pref{NLOa3}. For practical applications (fits to lattice data) there is no need to keep track of the individual LECs. However, one should make sure that the LEC combinations in the final results for observables are linearly independent, and this is the reason why we quoted all terms contributing to ${\cal L}_{\rm NLO}$.

Finally, the last NLO correction to the pion masses stems from the LO results. Setting $t=Xa$ in \pref{charMpi} and \pref{neutMpi} we find a O$(M_{\pm}^2a^2)$ correction to the charged pion mass and O($M_{0}^2a^2,a^4)$ corrections to the neutral pion mass. These are NLO corrections and need to be taken into account as well.

\subsection{Pion masses to NLO}\label{sect:NLOpionmass}

With the results given in the previous section the NLO calculation of the pion masses is straightforward. The 1-loop diagrams contributing to the self energy of the pions are all tadpole diagrams and stem from \pref{VertexLkin} and \pref{Verttotal}. Except for the vertex from the last term in \pref{Verttotal} these vertices are just the ones that also contribute in a continuum ChPT calculation. 
The only (but crucial) difference is that one has to keep track of the flavor indices for the pion in the loop, because the charged and the neutral pion have different masses. 
 All loop diagrams lead to the familiar scalar integrals, which are conveniently regularized by dimensional regularization. The divergencies are removed by introducing renormalized LECs at NLO.\footnote{Besides the divergence proportional to $\epsilon^{-1}$ the finite part $\ln 4\pi - \gamma +1$ is also subtracted.}
For the charged pion mass we find the result
\begin{align}\label{charMpiNLO1}
M^2_{\pi^\pm,{\rm NLO}} &= M^2_{\pm}\Bigg(1+\frac{M_0^2}{32\pi^2f^2}\ln\frac{M_0^2}{\tilde{\mu}^2} - \frac{16}{f^2}M^2_{\pm}\bigg(L_{45} - 2 L_{68}\bigg) + C_1 a^2\Bigg)\,.
\end{align}
$M_{\pm}$ and $M_{0}$ are the LO masses in \pref{Propchar}, \pref{Propneut} and $\tilde{\mu}$ denotes the renormalization scale. Here the NLO LECs are renormalized couplings and depend on that scale, $L_{45}^r(\tilde{\mu}),L_{68}^r(\tilde{\mu})$ and $C_1^r(\tilde{\mu})$, but for brevity we drop the superscript and do not make the dependence on $\tilde{\mu}$ explicit.

As already announced in the introduction, the chiral log in \pref{charMpiNLO1} contains the neutral pion mass and not the charged one. If the mass splitting is large this is a non-negligible modification. However, note that we recover the correct continuum result if $a$ goes to zero, since $M_0 \rightarrow M_{\pm}$ in this limit.\footnote{That the charged pion contains a chiral log involving the neutral pion mass has already been noted in ref.\ \cite{Ueda:2008zz}. However, there is a discrepancy in the prefactor of the chiral log. The source of this discrepancy is currently looked for \cite{Aokiprivcom}. Note, however, that the result in ref.\ \cite{Ueda:2008zz} does not reproduce the correct continuum chiral log in the continuum limit.}

The LEC $C_1$ is a combination of LECs and incorporates the O($M_{\pm}^2a^2$) corrections to the charged pion mass from various sources, as discussed at the end of the last section. It is easily checked that ${\cal L}_{a^4}$ in \pref{NLOa4} does not contribute to the charged pion mass, so there is no O($a^4$) shift in \pref{charMpiNLO1}. This is also expected qualitatively. Suppose $c_2>0$. In this case there exists an Aoki phase at $\mu=0$, and the charged pion mass must vanish for $\mu\rightarrow 0$. This excludes an O($a^4$) correction in the result for the charged pion mass. In fact, for the same reason there is no O($a^2$) shift in the LO mass in \pref{charMpi}.

In practical applications it is convenient to  introduce $\Lambda_3$, defined by
\begin{align}
16(L_{45} - 2L_{68}) &= \frac{1}{32\pi^2} \ln \frac{\Lambda_3^2}{\tilde{\mu}^2}.
\end{align}
In terms of $\Lambda_3$ we can rewrite \pref{charMpiNLO1} according to
\begin{align}\label{charMpiNLO2}
M^2_{\pi^\pm,{\rm NLO}} &= M^2_{\pm}\Big(1+\frac{M_0^2}{32\pi^2f^2}\ln\frac{M_0^2}{\Lambda_3^2} + C_{M_{\pm}} a^2\Big)\,.
\end{align}
The new combination of LECs, $C_{M_{\pm}}$, involves $C_1$ and $L_{45} - 2 L_{68}$. The scale dependence drops out in this combination and $C_{M_{\pm}}$ is scale independent.

The calculation of the neutral pion mass is completely analogous, and we find
\begin{align}
M^2_{\pi^0,{\rm NLO}} &= M^2_{\pm}\Bigg(1+\frac{1}{32\pi^2f^2}\bigg(2M^2_{\pm}\ln\frac{M_{\pm}^2}{\tilde{\mu}^2} - M_0^2\ln \frac{M_{0}^2}{\tilde{\mu}^2} \bigg) - \frac{16}{f^2}M_{\pm}^2(L_{45} - 2 L_{68}) + C_2 a^2 \Bigg)\nn\\
& \phantom{=} \, +2c_2a^2\Bigg(1 - \frac{M_0^2}{8\pi^2f^2} \ln \frac{M_0^2}{\tilde{\mu}^2} + C_3 a^2\Bigg)\label{neutMpiNLO1}
\,.
\end{align}
This result contains both types of chiral logs, involving the charged and the neutral pion mass. Still, in the continuum limit $a\rightarrow 0$ we recover the continuum result, as expected. 

$C_2$ and $C_3$ are combinations of LECs associated with the O($M_{\pm}^2a^2)$ and $O(a^4$) corrections. As before, these renormalized coefficients are scale dependent and the superscript ``{\em r}'' is suppressed. As for the charged pion mass, these coefficients represent various NLO corrections stemming from \pref{NLOp2a2} - \pref{NLOa3} as well as from the corrections to the LO pion mass in \pref{neutMpi} with $t={\rm O}(a)$. In contrast to the charged pion mass there is a correction of O($a^4$). 

It is again convenient to replace $L_{45} - 2 L_{68}$ by $\Lambda_3$. Similarly, $C_2$ and $C_3$ can be traded for a dimensionful scale $\Xi_3$ and a dimensionless and scale invariant coefficient $C_{\!M_0}$:
\begin{align}
M^2_{\pi^0,{\rm NLO}} &= M^2_{\pm}\Bigg(1+\frac{1}{32\pi^2f^2}\bigg(2M^2_{\pm}\ln\frac{M_{\pm}^2}{\Lambda_3^2} - M_0^2\ln \frac{M_{0}^2}{\Lambda_3^2} \bigg) \Bigg)\nn\\
& \phantom{=} \, +2c_2a^2\bigg(1 - \frac{M_0^2}{8\pi^2f^2} \ln \frac{M_0^2}{\Xi_3^2} + C_{\!M_0} a^2\bigg)\label{neutMpiNLO2}
\,.
\end{align} 
For completeness we also quote the NLO result for the mass splitting. 
Although one could directly take the difference of \pref{neutMpiNLO2} and \pref{charMpiNLO2} it seems beneficial to start with the differences of \pref{neutMpiNLO1} and \pref{charMpiNLO1}. The contribution proportional to $L_{45} - 2 L_{68}$ drops out and the two O$(M_{\pm}^2a^2$) contributions can be combined at this stage. Then one proceeds as for the neutral pion mass and finds
\begin{align}
\Delta M^2_{\pi,{\rm NLO}} &=   2c_2a^2\bigg(1 - \frac{M_0^2}{8\pi^2f^2} \ln \frac{M_0^2}{\tilde{\Xi}_3^2} + C_{\!\Delta} a^2\bigg) \nn\\
&\phantom{=}  +\frac{M_{\pm}^2}{16\pi^2f^2}\bigg(M^2_{\pm}\ln\frac{M_{\pm}^2}{\Lambda_3^2} - M_0^2\ln \frac{M_{0}^2}{\Lambda_3^2} \bigg)\,.
\end{align}
$\tilde{\Xi}_3$ and $C_{\!\Delta}$ differ from the analogous coefficients in \pref{neutMpiNLO2} by terms proportional to $C_1$.

\subsection{Decay constant to NLO}

An equally important observable besides the pion mass is the pion decay constant $f_{\pi}$, defined by the one-pion matrix element of the axial vector current. In this section we compute  the decay constant $f_{\pi^{\pm}}$ of the charged pions to NLO. For simplicity we write $f_{\pi^{\pm}}=f_{\pi}$ since we never encounter the decay constant of the neutral pion. We will find that the chiral log in $f_{\pi}$ is also modified by a large pion mass splitting.

In twisted mass QCD one usually does not compute $f_{\pi}$ by the matrix element involving the axial vector current. Instead, one makes use of the so-called {\em indirect method} \cite{Frezzotti:2001ea,Jansen:2003ir} where the decay constant is given by 
\begin{align}
f_{\pi} & =  \frac{2\mu}{M^2_{\pi^{\pm}}}  G_{\pi} \,,\label{fpiIndirect1}\\
G_{\pi} &= |\langle 0| P^{a}(0)|\pi^{a}(\vec{p})\rangle|\,,\quad a\,=\,1,2 \,,\label{fpiIndirect2}
\end{align}
where $P^a(x)$ is the pseudo scalar density. Relation \pref{fpiIndirect1} is based on the exact Ward identity \cite{Frezzotti:2000nk}
\begin{align}\label{PCVCrelation}
\partial_{\mu}V_{\mu}^a &= -2\mu \epsilon^{3ab}P^b
\end{align}
involving the vector current $V_{\mu}^a$. At maximal twist the vector current corresponds to the physical axial vector current. This implies \pref{fpiIndirect1}, and its benefit is that one does not need to compute any renormalization factors like $Z_{\rm P}$ or $Z_{\rm A}$.

The right hand side of \pref{fpiIndirect1} is straightforwardly calculated in WChPT. The pion mass is given in the last section, and the missing piece is the matrix element $G_{\pi}$.
The expression for the pseudo scalar density can be found in ref.\ \cite{Sharpe:2004ny}. To LO it is the familiar expression from continuum ChPT,
\begin{align}
P^a_{\rm LO} & =  \frac{f^2B}{4}\langle \sigma_a(\Sigma-\Sigma^\dagger)\rangle. \label{LOPa}
\end{align}
Expanding to linear order in the pion fields we find the tree level result $G_{\pi,{\rm LO}}=fB$. 
Using the tree level result \pref{charMpi} for the charged pion mass we obtain
\begin{align}\label{fpiLO}
f_{\pi,{\rm LO}} &= f\sqrt{1-t^2}.
\end{align}
This is the result for the right hand side of \pref{fpiIndirect1} for an arbitrary twist angle.\footnote{The result correctly vanishes for $t=0$. This corresponds to the untwisted case. The Ward identity \pref{PCVCrelation} still holds (for $\mu=0$), however, $V_{\mu}$ is the physical vector current and its one-pion matrix element vanishes.} At maximal twist ($t=0$) it turns into the well-known LO result for the decay constant. If $t={\rm O}(a)$ it receives, after expanding the square root, an O($a^2$) correction. 
Note that the origin of the factor $\sqrt{1-t^2}$ is the result for the pion mass. $G_{\pi,{\rm LO}}$ is exactly as in continuum ChPT and does not add any modifications due to the non-trivial ground state.

In order to compute the decay constant to NLO we need the NLO expression for the pseudo scalar density in the LCE regime. Most terms can be taken from ref.\ \cite{Sharpe:2004ny} where $P^a_{\rm NLO}$ has been derived for the GSM regime. Missing are the O($a^2$) corrections, but these are easily constructed (see below).

The general structure of $P^a$ at NLO is
\begin{align}
P^a_{\rm NLO} & =  P^a_{\rm LO} (1 + \Delta P_{\rm NLO, GSM}+\Delta P_{a^2}). \label{NLOPa}
\end{align}
The correction $\Delta P_{\rm NLO, GSM}$ can be found in eq.\ (47) of \cite{Sharpe:2004ny} (where it is denoted by ${\cal D}$)\footnote{The LEC $W$ in ref.\ \cite{Sharpe:2004ny} corresponds to $W_{68}$ in our case since we directly started with the chiral lagrangian parameterized in terms of the shifted mass.},
\begin{align}\label{DeltaPa}
\Delta P^a_{\rm NLO} &= -\frac{4L_{45}}{f^2}\langle \partial_{\mu}\Sigma \partial_{\mu}\Sigma^{\dagger}\rangle + \frac{8L_{68}}{f^2} \langle \hat{M}^{\dagger}\Sigma + \Sigma^{\dagger}\hat{M}\rangle +\frac{4\hat{a}W_{68}}{f^2}\langle\Sigma+\Sigma^{\dagger} \rangle\,.
\end{align}
These terms are the corrections of O($p^2,M,a)$ to the leading 1 in \pref{NLOPa}. The correction proportional to $L_{45}$ contributes at NNLO only and can be ignored in the following. In the mass term we can set the untwisted mass $m$ to zero since we are interested in the case of maximal twist only. The correction proportional to $\hat{a}W_{68}$ is effectively a O$(a^2)$ contribution at maximal twist because of the contribution $\langle\Sigma+\Sigma^{\dagger} \rangle$ (recall the discussion after eq.\ \pref{MEleadingtot}). 

The method used in ref.\ \cite{Sharpe:2004ny} for the construction of the pseudo scalar density is easily extended to the O($a^2)$ terms in $\Delta P_{a^2}$. The pseudo scalar density is obtained by a functional derivative of the effective action with respect to the mass, which is promoted to a space-time dependent spurion field in an intermediate step. For example, the O($a$) correction in \pref{DeltaPa} is obtained from the ${\cal L}_{Ma}$ lagrangian in \pref{NLOeffMa}. In complete analogy the lagrangian ${\cal L}_{Ma^2}$ in \pref{NLOMa2} yields the O($a^2$) corrections
\begin{align}
\Delta P_{a^2}&= -\frac{4a^2}{f^2}\Big(b_1 +b_2 \langle \Sigma+\Sigma^{\dagger}\rangle^2\Big)\,.
\end{align}
The term proportional to $b_2$ is effectively an O($a^4$) term because of the factor $\langle \Sigma+\Sigma^{\dagger}\rangle^2$. 

Dropping all terms that contribute beyond NLO only we can use the following (incomplete) NLO expression for the pseudo scalar density:
\begin{align}\label{effPaNLO}
P^a_{\rm NLO} & =  P^a_{\rm LO} \Bigg(1 + \frac{8L_{68}}{f^2} M^2_{\pm}\langle i\sigma_3( \Sigma^{\dagger} - \Sigma)\rangle -\frac{4}{f^2}b_{1,{\rm eff}} a^2 \Bigg)\,.
\end{align}
where we introduced the effective LEC $b_{1,{\rm eff}} = b_1 - 8W_{68}W_0 X$ which includes the remnant O($a^2$) correction from the O($a$) term in \pref{DeltaPa} (as before we have set $t=Xa$).\footnote{We remark that the expression of the pseudo scalar density is determined by the Ward identity \pref{PCVCrelation}. In lattice QCD many pseudo scalar densities can be defined, all differing by O($a$). The corresponding expressions in WChPT differ too, since they have to accommodate these differences. This is analogous to the vector and axial vector currents, which have been discussed in ref.\ \cite{Aoki:2009ri}.}

The 1-loop calculation of the matrix element $G_{\pi}$ is now standard. Expanding $P^a_{\rm LO}$ in \pref{effPaNLO} up to cubic order in pion fields one obtains the terms that lead to the 1-loop corrections. Since $P^a_{\rm LO}$ is the familiar continuum expression, the calculation is as in continuum ChPT, except for the fact that one has to keep track of the flavor index of the pion in the loop since the masses are different. 
The result of the calculation is
\begin{align}
G_{\pi,{\rm NLO}} &= fB\Bigg(1- \frac{M^2_{\pm}}{32\pi^2f^2}\ln\frac{M^2_{\pm}}{\tilde{\mu}^2} + \frac{8M_{\pm}^2}{f^2}(4L_{68} - L_{45}) + C_4a^2\Bigg)\,.\label{GpiNLO}
\end{align}
We introduced $C_4$ as a short hand notation for the contributing combination of LECs. It contains $b_{1,{\rm eff}}$ and also $a_1$ of the lagrangian ${\cal L}_{a^2p^2}$ (it contributes to the wave function renormalization). Forming the ratio in \pref{fpiIndirect1} with the result \pref{charMpiNLO1} for the charged pion mass we finally obtain the NLO result for the decay constant:
\begin{align}
f_{\pi,{\rm NLO}} &= f\Bigg(1- \frac{1}{32\pi^2f^2}\bigg(M^2_{\pm}\ln\frac{M^2_{\pm}}{\tilde{\mu}^2}+ M^2_{0}\ln\frac{M^2_{0}}{\tilde{\mu}^2}\bigg) + \frac{8M_{\pm}^2}{f^2} L_{45} + (C_4-C_1)a^2\Bigg)\,.\label{fpiNLO}
\end{align}
Both \pref{GpiNLO} and \pref{fpiNLO} turn into the known continuum ChPT result for $a\rightarrow 0$. Also, the GSM result at maximal twist \cite{Sharpe:2004ny} is correctly reproduced. 
The analytic lattice spacing dependence is O($a^2$), as expected from automatic O($a$) improvement. However, the chiral log is not the naively expected one: the averaged chiral log $[M^2_{\pm}\ln M^2_{\pm} + M^2_{0}\ln M^2_{0}]/2$ appears in the result.\footnote{This is also reminiscent of the 1-loop result in staggered ChPT: The pion decay constant involves the average of all sixteen chiral logs that one can form with the available taste partners \cite{Aubin:2003uc}.}

As for the pion masses  it is convenient to slightly rewrite the result \pref{fpiNLO}. In terms of the scale $\Lambda_4$, defined by
\begin{align}
8L_{45} &= \frac{1}{16\pi^2} \ln \frac{\Lambda_4^2}{\tilde{\mu}^2},
\end{align}
the result for the decay constant turns into
\begin{align}\label{fpiNLOfinal}
f_{\pi,{\rm NLO}} &= f\Bigg(1- \frac{1}{32\pi^2f^2}\bigg(M^2_{\pm}\ln\frac{M^2_{\pm}}{\Lambda_4^2}+ M^2_{0}\ln\frac{M^2_{0}}{\Lambda_4^2}\bigg)+ C_{\!f}a^2\Bigg)\,.
\end{align}
In analogy to $C_{M_{\pm}}$ we introduced the notation $C_{\!f}$ for the final combination of  O($a^2$) LECs.

We followed the actual numerical computation of the decay constant and calculated $f_{\pi}$ with the indirect method. Alternatively, one can also compute the standard matrix element involving the (physical) axial vector current. The result must be the same, of course. We partially confirmed this by calculating the 1-loop correction for the axial vector current matrix element.  We indeed found the chiral logs as in \pref{fpiNLO}. This also provided a non-trivial check for the pion mass computation in section \ref{sect:NLOpionmass}.

\subsection{Finite volume corrections}\label{ssect:FV}

So far all the calculations were done in infinite volume, but the finite volume (FV) corrections are known \cite{Gasser:1986vb,Gasser:1987ah,Gasser:1987zq} and easily included. Most relevant in practice is a finite space-time volume with geometry $L^3\times T$ and periodic boundary conditions in each direction. As has been shown in ref.\ \cite{Gasser:1987zq}, the chiral lagrangian is as in infinite volume and the finite volume dependence enters through the pion propagators only.
The position space propagator, obtained by Fourier transforming \pref{Propchar} and \pref{Propneut}, involves a sum over the discrete pion momenta instead of an integration.  

As a concrete example we quote here the modifications of the previously derived results for a finite spatial volume $L^3$, assuming the temporal extend $T$ to be much larger so that it can be taken infinite. In this case the FV corrections are included by the simple replacement \cite{Bernard:2001yj}\footnote{Ref.\ \cite{Bernard:2001yj} denotes the FV correction by $\delta_1$. The notation $\tilde{g}_1$ is used in refs.\ \cite{Colangelo:2003hf,Colangelo:2005gd}.}
\begin{align}
\ln \frac{M^2}{\Lambda^2} \,\,\rightarrow\,\, \ln \frac{M^2}{\Lambda^2} + \tilde{g}_1(ML)\,,
\end{align}
\begin{align}
\tilde{g}_1(ML)&= \frac{4}{ML}\sum_{\vec{n}\neq 0} \frac{K_1(|\vec{n}|ML)}{|\vec{n}|}\,,\quad\qquad |\vec{n}|\,=\sqrt{n_1^2+n_2^2+n_3^2}\,,\label{Defg1tilde}
\end{align}
in every chiral log. $K_1$ is the modified Bessel function of the second kind. The sum runs over all triples $\vec{n}=(n_1,n_2,n_3)$ where the $n_k$ are integers. Taking into account the degeneracies in this three-dimensional sum it collapses to a simple one-dimensional sum \cite{Colangelo:2003hf,Colangelo:2005gd}. The Bessel function $K_1$ is exponentially small for large arguments, $K_1(ML) \approx \exp(-ML)/\sqrt{ML}$, hence the sum in \pref{Defg1tilde} converges rapidly and the FV correction $\tilde{g}_1(ML)$ is small. 

The results of the previous section contain chiral logs with the charged and the neutral pion mass. We therefore obtain FV corrections governed by the arguments $M_{\pm}L$ and $M_0L$. The latter are significantly (exponentially) larger if the neutral pion mass is much smaller than the charged one, as has been pointed out in ref.\ \cite{Colangelo:2010cu}.

\section{Numerical analysis}

The calculations of the previous section were triggered by the numerical results of the ETM collaboration in ref.\ \cite{Baron:2010bv}. As already mentioned in the introduction, measurements of the charged and neutral pion masses showed that the latter is significantly lighter, with $M_{\pi^0}/M_{\pi^{\pm}}$ approximately  $0.54$ and $0.77$ for two of the simulated quark masses at $a\approx 0.078$fm. These values correspond to roughly 0.71 and 0.41 for the ratio
\begin{align}
\frac{M^2_{\pi^{\pm}} - M^2_{\pi^{0}}}{M^2_{\pi^{\pm}}}&\approx \frac{|2c_2a^2|}{2B\mu},
\end{align}
and this tells us that at least these data points are in the LCE regime. 
It is therefore interesting to check whether the data show evidence for the modified chiral logs and, provided this is the case, whether the extraction of  Gasser-Leutwyler coefficients is affected by using the results of the previous section instead of the continuum ChPT results.\footnote{A similar analysis of quenched lattice data can be found in ref.\ \cite{Aoki:2006js}.}

Data for the charged pion mass and the pion decay constant is given in table 8 of ref. \cite{Baron:2010bv}. Measurements have been made for two different lattice spacings ($a\approx$ 0.08585fm and  0.0782fm) and for various (6 and 5) different light quark masses. The charged pion mass ranges roughly between 270 and 510MeV. In total there exist 22 data points for a combined fit to the data. This is rather large compared to the number of fit parameters in the expressions \pref{charMpiNLO2} and \pref{fpiNLOfinal}: Four continuum ChPT parameters $f,B,\Lambda_3,\Lambda_4$, and three more associated with the nonzero lattice spacing, $c_2,C_{M_{\!\pm}}$ and $C_{\!f}$. Even if one drops a few data points for the heavier pion masses there are still enough data points to perform a fit.

All fit results presented in this section were obtained by fits to the dimensionless lattice data for $af_{\pi}$ and the ratio 
\begin{align}
R&=\frac{(aM_{\pi^{\pm}})^2}{a\mu_0}\,,
\end{align}
where $\mu_0$ denotes the bare twisted quark mass.\footnote{The conventions in ref.\ \cite{Baron:2010bv} are such that $f\approx$ 130MeV. Therefore, in the results of the previous section the replacement $f^2\rightarrow f^2/2$ in the prefactor of the chiral logs has to be made.} The renormalized quark mass $\mu$ used in the previous section renormalizes multiplicatively, thus $\mu$ is proportional to $\mu_0$ and we have $B\mu=B_0\mu_0$. 
Instead of quoting the fit results for $\Lambda_3,\Lambda_4$ we give the results for 
\begin{align}
\overline{l}_{3,4} &= \ln\Bigg(\frac{\Lambda_{3,4}^2}{M_{\pi,{\rm phys}}^2}\Bigg)\,,
\end{align}
where $M_{\pi,{\rm phys}}=139.6$MeV denotes the physical pion mass.  

Results of fits to the data with the pion mass splitting ignored in the chiral logs have been reported in ref.\ \cite{Baron:2010bv}. We also give results for such fits, but the results are expected to differ slightly for two reasons. In ref. \cite{Baron:2010bv} the resummed FV corrections of \cite{Colangelo:2005gd} were used, which differ from the FV formulae of section \ref{ssect:FV} even if we set $c_2$ equal to zero.\footnote{Note that the formulae of \cite{Colangelo:2005gd} contain two more NLO LECs, $\overline{l}_1$ and $\overline{l}_2$.} 
Another difference concerns the error analysis. The statistical errors for the fit parameters in ref.\ \cite{Baron:2010bv} are estimated by generating bootstrap samples from the bare correlation functions for the pion mass and decay constant. This cannot be done having only the data in table 8 of \cite{Baron:2010bv} available. Instead, the fit results and error estimates given here were obtained by fitting the data with the Levenberg-Marquardt algorithm \cite{NumRec}.

\begin{table}[t]
\begin{tabular}[t]{l | cc | cc | cc |}\hline\hline
Fit ($\beta =1.90$)& \multicolumn{2}{c|}{I}& \multicolumn{2}{c|}{II}& \multicolumn{2}{c|}{III}\\ 
Fit range: $\;a\mu_{0,\rm min}$ & \multicolumn{2}{c|}{ $0.003$} &\multicolumn{2}{c|}{ $0.004$} & \multicolumn{2}{c|}{ $0.003$}\\
\phantom{Fit range:} $\;a\mu_{0,\rm max}$ & \multicolumn{2}{c|}{ $0.01$} &\multicolumn{2}{c|}{ $0.01$} & \multicolumn{2}{c|}{ $0.008$}\\
maximal $M_{\pi^{\pm}}$ (MeV)& \multicolumn{2}{c|}{ 512} &\multicolumn{2}{c|}{ 512} & \multicolumn{2}{c|}{456}\\
\hline
$2B_0a$ & 5.49(4) & 5.24(7) & 5.45(5) & 5.24(18) & 5.52(5) & 5.24(9)\\
$f$ (MeV) &  $119.9(1.0)$ & $129.5(2.9)$ & 120.5(1.1) & 129.4(8.2) & 119.8(1.2) & 129.7(3.3)\\
$\lt$ & $3.47(7)$ & $3.09(17)$ & 3.41(9) & 3.08(37)&3.52(11)& 3.11(24)\\
$\lf$ & $4.74(2)$& $4.71(4)$& 4.74(3) & 4.71(7) & 4.74(4)& 4.70(6)\\ 
$-2c_2a^2$ (MeV$^2$)\hspace{4mm} & - & $[214(27)]^2$ & - & $[213(90)]^2$& - &$[215(29)]^2$ \\ \hline
$n_{\rm data}$ & 12 & 12& 10 & 10 & 10 & 10 \\
$\chi^2/n_{\rm dof}$ & 8.81/8 & 5.48/7& 6.50/6& 5.46/5& 8.36/6 &5.34/5 \\
$Q$ & 0.55 & 0.86 & 0.59 & 0.71 & 0.40 & 0.72\\
\hline\hline
\end{tabular}
\vspace{1cm}

\begin{tabular}[t]{l | cc | cc | cc |}\hline\hline
Fit ($\beta =1.95$)& \multicolumn{2}{c|}{I}& \multicolumn{2}{c|}{II}& \multicolumn{2}{c|}{III}\\ 
Fit range: $\;a\mu_{0,\rm min}$ & \multicolumn{2}{c|}{ $0.0025$} &\multicolumn{2}{c|}{ $0.0035$} & \multicolumn{2}{c|}{ $0.0025$}\\
\phantom{Fit range:} $\;a\mu_{0,\rm max}$ & \multicolumn{2}{c|}{ $0.0085$} &\multicolumn{2}{c|}{ $0.0085$} & \multicolumn{2}{c|}{ $0.0075$}\\
maximal $M_{\pi^{\pm}}$ (MeV)& \multicolumn{2}{c|}{496} &\multicolumn{2}{c|}{496} & \multicolumn{2}{c|}{461}\\
\hline
$2B_0a$ & 4.94(4) & 4.72(8) & 4.96(6) & 4.61(8) & 4.98(5) & 4.83(18)\\
$f$ (MeV) &  $119.9(1.3)$ & $128.0(3.3)$ & 120.5(1.4) & 134.3(3.5) & 119.3(1.3) & 124.3(6.4)\\
$\lt$ & $3.66(8)$ & $3.36(18)$ & 3.70(10) & 3.16(23)&3.74(9)& 3.60(24)\\
$\lf$ & $4.67(3)$& $4.62(5)$& 4.65(3) & 4.52(7) & 4.68(3)& 4.66(6)\\ 
$-2c_2a^2$ (MeV$^2$)\hspace{4mm} & - & $[190(31)]^2$ & - & $[249(22)]^2$& - &$[150(89)]^2$ \\
\hline
$n_{\rm data}$ & 10 & 10& 8 & 8 & 8 & 8  \\
$\chi^2/n_{\rm dof}$ & 13.3/6 & 10.6/5& 11.2/4& 6.3/3& 4.1/4 &3.80/3 \\
$Q$ & 0.10 & 0.23 & 0.08 & 0.39 & 0.67 & 0.70\\
\hline\hline
\end{tabular}
\caption{\label{fig:sepFit} Results for the fits to the data at fixed lattice spacings, $a\approx 0.086$fm ($\beta =1.9$, top) and $a\approx 0.078$fm ($\beta=1.95$, bottom). Right subcolumns correspond to fits with $c_2$ as a free fit parameter, left subcolumns to fits with $c_2$ set to zero. The parameters $C_{M_{\pm}}$, $C_{\! f}$ are always set to zero in these fits (see text).}
\end{table}

\begin{table}[bth]
\begin{tabular}[t]{l | c | c | c |}
 &   $\beta=1.90$ &  $\beta=1.95$ & combined\\ \hline
$f$ (MeV) &  $120.956(70)$ & $121.144(83)$ & 121.031(54)\\
$\lt$ & $3.435(61)$ & $3.698(73)$ & 3.537(47)\\
$\lf$ & $4.773(21)$& $4.673(25)$& 4.735(17) \\
\hline
\end{tabular}
\caption{\label{fig:ETMCFit} Selected results of the fits performed by the ETM collaboration, taken from table 9 of Ref.\ \cite{Baron:2010bv}. The first two columns refer to the separate fits, the last one to the combined fit.}
\end{table}

In order to check for these potential differences we performed separate fits to the data at fixed lattice spacing with $C_{M_{\pm}}$ and  $C_{{\!f}}$ set to zero.\footnote{For fits at one lattice spacing the constants $B_0$ ($f$) and  $C_{M_{\pm}}$ ($C_{\! f}$) would not be independent if the chiral log was absent. Even in the presence of the (small) chiral log these fit parameters are not very well determined individually.}
The results are given in table \ref{fig:sepFit}. We performed three fits that differ in the data points included in the fit. Fit I included all data points while for fit II (III) the data points at the smallest (heaviest) quark mass were excluded. For all fit ranges two fits were done, one that includes $c_2$ as a fit parameter (right subcolumn) and one without, setting $c_2$ equal to zero (left subcolumn, indicated by a dash).

Fit II with $c_2=0$ has been done by the ETM collaboration and their results in table 9 of \cite{Baron:2010bv} should be compared with ours (for the readers convenience we have collected the relevant fit results in table \ref{fig:ETMCFit}). The data points for the smallest quark mass were dropped by the ETM collaboration because the data did not fully comply with the tuning condition $m_{\rm PCAC}/\mu_0 < 0.1$ for maximal twist (see section 3.2 in ref.\ \cite{Baron:2010bv}).

The mean values for $f$ and the LECs $\overline{l}_{3,4}$ agree to a very good degree. Our errors for $\overline{l}_{3,4}$ are somewhat larger, which is not unexpected. The error for the decay constant, on the other hand, is slightly puzzling: Our errors for $f$ are roughly at the one percent level, while the error in ref.\ \cite{Baron:2010bv} is more than an order of magnitude smaller. The reason for this strong discrepancy is not clear to us.

Let us turn to the fits in  table \ref{fig:sepFit} that include $c_2$ as a fit parameter. Qualitatively we can say that the data sets prefer a negative value for $c_2$ with $-2c_2a^2\approx (200{\rm MeV})^2$. Although the error is quite large the sign is in agreement with a neutral pion lighter than the charged ones. All fits with $c_2$ as a fit parameter have slightly better values for the $\chi^2/n_{\rm dof}$ and the goodness of the fit $Q$, but the improvement is not dramatic.
Noteworthy is that the values for $f$ are systematically larger, the ones for $\overline{l}_3$ systematically smaller compared to the fits with $c_2$ set to zero. 
The worst discrepancy with a $3.9\sigma$ difference appears  in Fit II at $\beta=1.95$ for $f$, but most differences are (roughly) between $1\sigma$ and $2.5\sigma$. These differences are sizable and not negligible. However, before one can draw firm conclusions these results need to be corroborated by fits that properly take into account any correlations in the data.

\begin{table}[t]
\begin{tabular}[t]{l | cc | cc | cc |}\hline\hline
Fit (both $\beta$ values)& \multicolumn{2}{c|}{I}& \multicolumn{2}{c|}{II}& \multicolumn{2}{c|}{III}\\ 
Fit range: $\;a\mu_{0,\rm min}$ & \multicolumn{2}{c|}{ $0.0025$} &\multicolumn{2}{c|}{ $0.0035$} & \multicolumn{2}{c|}{ $0.0025$}\\
\phantom{Fit range:} $\;a\mu_{0,\rm max}$ & \multicolumn{2}{c|}{ $0.01$} &\multicolumn{2}{c|}{ $0.01$} & \multicolumn{2}{c|}{ $0.008$}\\
maximal $M_{\pi^{\pm}}$ (MeV)& \multicolumn{2}{c|}{ 512} &\multicolumn{2}{c|}{ 512} & \multicolumn{2}{c|}{456}\\
\hline
$2B_0a$ & 4.57(11) & 4.39(11) &4.52(12) & 4.37(12) & 4.69(13) & 4.54(14) \\
$f$ (MeV) &  $111.3(2.2)$ & $116.2(2.5)$ & 112.2(2.3) & 116.8(2.7) & 112.4(2.4) & 116.2(3.0)\\
$\lt$ & $3.44(7)$ & $3.09(13)$ & 3.40(8) & 2.98(19)&3.60(8)& 3.34(17)\\
$\lf$ & $4.69(4)$& $4.62(5)$& 4.69(4) & 4.56(7) & 4.70(5)& 4.63(6)\\ 
$-2c_2a^2$ (MeV$^2$)\hspace{4mm} & - & $[187(19)]^2$ & - & $[216(28)]^2$& - &$[171(34)]^2$ \\
$C_{M_{\pm}}a^2$ & 0.19(2) & 0.19(3)& 0.20(3)& 0.19(3)& 0.17(3)& 0.17(3)  \\
$C_{\!f}a^2$ & 0.10(2) &  0.13(2)& 0.10(2)& 0.14(2)& 0.09(2)& 0.11(2)\\ \hline
$n_{\rm data}$ & 22 & 22& 18 & 18 & 18 & 18 \\
$\chi^2/n_{\rm dof}$ & 27.6/16 & 20.7/15& 23.8/12& 18.8/11& 14.1/12 &11.7/11 \\
$Q$ & 0.12 & 0.42 & 0.09 & 0.28 & 0.59 & 0.76\\
\hline\hline
Fit (both $\beta$ values)& \multicolumn{2}{c|}{IV}& \multicolumn{2}{c|}{V}& \multicolumn{2}{c|}{VI}\\ 
Fit range: $\;a\mu_{0,\rm min}$ & \multicolumn{2}{c|}{ $0.0035$} &\multicolumn{2}{c|}{ $0.0025$} & \multicolumn{2}{c|}{ $0.0025$}\\
\phantom{Fit range:} $\;a\mu_{0,\rm max}$ & \multicolumn{2}{c|}{ $0.008$} &\multicolumn{2}{c|}{ $0.006$} & \multicolumn{2}{c|}{ $0.005$}\\
maximal $M_{\pi^{\pm}}$ (MeV)& \multicolumn{2}{c|}{456} &\multicolumn{2}{c|}{397} & \multicolumn{2}{c|}{363}\\
\hline
$2B_0a$ & 4.70(15) &4.57(16) & 4.77(16) & 4.59(17) & 4.93(21) & 4.71(25) \\
$f$ (MeV) &  $113.1(2.6)$ & $116.8(3.4)$ & 114.1(2.7) & 119.2(3.5) & 112.6(3.3) & 118.3(5.3)\\
$\lt$ & $3.62(10)$ & $3.33(25)$ & 3.65(16) & 3.21(31)&3.40(31)& 2.65(89)\\
$\lf$ & $4.68(5)$& $4.58(10)$& 4.74(7) & 4.61(11) & 4.85(15)& 4.61(33)\\ 
$-2c_2a^2$ (MeV$^2$)\hspace{4mm} & - & $[197(55)]^2$ & - & $[189(31)]^2$& - &$[195(55)]^2$ \\
$C_{M_{\pm}}a^2$ & 0.17(3) & 0.16(3)& 0.15(3)& 0.15(3)& 0.11(5)& 0.10(5)  \\
$C_{\!f}a^2$ & 0.09(2) &  0.12(3)& 0.07(2)& 0.09(2)& 0.06(3)& 0.10(4)\\ \hline
$n_{\rm data}$ & 14 & 14& 14 & 14 & 10 & 10  \\
$\chi^2/n_{\rm dof}$ & 11.1/8 & 9.70/7& 8.7/8& 5.96/7& 3.1/4 &2.68/3 \\
$Q$ & 0.52 & 0.64 & 0.73 & 0.92 & 0.93 & 0.95\\
\hline\hline
\end{tabular}
\caption{\label{fig:combFit} Results for the combined fits to the data for both lattice spacings. 
As in table \ref{fig:sepFit}, right subcolumns correspond to fits with $c_2$ as a free fit parameter, left subcolumns to fits with $c_2$ set to zero.
The values for the fit parameters involving the lattice spacing refer to $a\approx0.086$ fm ($\beta=1.9$).}
\end{table}

The main motivation for the WChPT calculations in the previous section is a combined fit to the data at both lattice spacings, since this amounts in a combined chiral and continuum extrapolation. The results of such fits are given in table \ref{fig:combFit}.
As before, various fits were done with different ranges for the pion masses.
Fits I to III include the same data points as in the separate fits. Fits with even more data points excluded were done as well: The smallest and the largest pion mass (at each lattice spacing) were dropped in fit IV, while the largest two (three) pion mass data points were excluded in fit V (VI). All fits include $C_{M_{\pm}}$ and $C_{\! f}$ as free fit parameters. 
The ratio of the two lattice spacings, on the other hand, is not a fit parameter but included as the fixed ratio $r_a= 0.0782/0.08585$.
Results in table \ref{fig:combFit} that include the lattice spacing refer to the values at the larger lattice spacing $a\approx 0.08585$fm ($\beta=1.9$). 

A combined fit requires the ratio of quark mass renormalization factors $Z_{\mu}$ at the two lattice spacings. This ratio is not available to us so we set it to 1. We expect this to be a good approximation because the two lattice spacings are very close: The finer lattice spacing is less than 10 per cent smaller than the coarse one.\footnote{We also did fits with the renormalization factor ratio included as a free fit parameter. We obtained ratios between 0.98 and 1.02 with an error of about 0.02. The results for the other fit parameter agree with the ones given in table \ref{fig:combFit} within the errors. However, the error estimates for the fit parameters $2B_0a$ and $C_{M_{\pm}}a^2$ are larger by a factor 4 to 5.}

None of the results in table \ref{fig:combFit} can be compared with results in ref.\ \cite{Baron:2010bv}. Although the ETM collaboration has done a combined fit, the coefficients $C_{M_{\pm}}$ and $C_{\! f}$ were set to zero in this fit.   

From table \ref{fig:combFit} we draw the following conclusions.
\begin{enumerate}
\item Quite generally, all fits are satisfactory with respect to their $\chi^2$ and $Q$ values, even the ones with all data points included. 
The largest values for $\chi^2/n_{\rm dof}$ is about 2 in fit II. 
The quality of the fits improve if data points are dropped. Comparisons of the fits I with II and III as well as II with IV shows that $\chi^2$ decreases substantially if the data for the heaviest pion mass are excluded. The improvement is less significant for dropping the smallest pion mass data.  This is in agreement with ChPT as a low-energy effective theory. 

As already mentioned, the reason for dropping the data at the smallest pion mass by the ETM collaboration was the potential violation of tuning to maximal twist. It is not easy to exactly quantify a small mistuning, but the results in table \ref{fig:combFit} show that including the data at the lightest pion mass is less influential than including the data at the heaviest pion mass. The central values for fits IV and V are in good agreement and either of it seems (at least to us) to be a good candidate for obtaining reliable fit results. 

\item Fits with $c_2$ as a fit parameter are always better than without. However, the improvement is less significant for the fits with the heavier mass data excluded. 
Nevertheless, the data prefer a negative $c_2$ in agreement with a neutral pion mass smaller than the charged one. The LO pion mass splitting $-2c_2a^2$ is roughly $(200$MeV)$^2$ (at $a\approx 0.086$fm) with a large error. Note that the errors given in table \ref{fig:combFit} are slightly misleading: The square root of the fit parameter $-2c_2a^2$ is quoted and its relative error is half as big. 

Note that the neutral pion mass $M_{{\pi^0},{\rm NLO}}$ is not predicted by the values in table \ref{fig:combFit} because it depends on two extra parameters, $\tilde{\Xi}_3$ and $C_{M_0}$. 
Hence, we cannot check whether our fit results are in agreement with the direct measurements of the neutral pion mass. Moreover, it is not possible to include the data for the neutral pion mass in the fit, since only two data points are given in ref.\ \cite{Baron:2010bv}. Once more data become available it will be very interesting to attempt combined fits with the neutral pion mass data included. 

\item Comparing the fit results with and without $c_2$ as a fit parameter one can observe: 
(i) the central values agree within errors, (ii) the central values for $f$ are somewhat larger, and smaller for the Gasser-Leutwyler coefficients $\overline{l}_{3,4}$, (iii) the errors are in general larger with $c_2$ included. 
 Whether these observations persist for smaller errors cannot be said here. 
An error analysis as the one in ref.\ \cite{Baron:2010bv} may lead to smaller errors and perhaps to  a firmer conclusion. 

\item Our last observation concerns the analytic O($a^2$) corrections to the LO LECs, $B_0$ and $f$. Consider, for example,  fit II with $c_2$ excluded from the fit. The results for $C_{M_{\pm}}$ and $C_{\! f}$ in the combined fit mean that there is a 20\% and 10\%  O($a^2$) error in $B_0$ and $f$ at $a\approx 0.086$fm. These numbers are consistent with the separate fit at this lattice spacing (fit II in top of table \ref{fig:sepFit}). Note that we have set $C_{M_{\pm}}$ and $C_{\! f}$
equal to zero in the separate fit, so the correction associated with these LECs is effectively absorbed in $B_0$ and $f$: $f$ in the separate fits corresponds to $f(1+a^2C_{\! f})$ in the combined fit, and analogously for $B_0$. 

Our results here show that the separate fits can be quite misleading for the estimates in the continuum limit. For example, the central value for $f$ is unchanged for the two separate fits II, and it might be tempting to interpret this as the O($a^2$) correction being very small and almost negligible. However, the central value for $f$ in the combined fit is almost 10\% smaller. 

We do not claim that the correct continuum limit can only be obtained with a combined fit.
However, a proper continuum extrapolation of the results from the separate fits is not possible yet since data at two lattice spacings only are available. Moreover, the two lattice spacings do not cover a wide range with the fine lattice spacing being only 9 percent smaller than the coarse one. More data at significantly smaller  lattice spacings seem necessary to shed more light on this issue.

\end{enumerate}

\begin{table}[t]
\begin{tabular}[t]{l | cclclclclcl |}\hline\hline
$a\mu_0$ & $M_0/M_{\pm}$&  &$ r_{\rm logs}$ & & $M_0 L$ & & $M_{\pm}L$ & & $r(M_0 L)$ & & $r(M_{\pm}L)$ \\ \hline
0.003 & 0.5367 && 0.5476 && 2.7498 && 3.7534 & & -0.288 & & -0.073 \\
0.004 &	0.6525 && 0.7706 & & 3.5011 & & 4.3341 & & -0.101 & & -0.036 \\
0.005 & 0.7220 & &0.8682  & &4.1175	& & 4.8456 & & -0.047 & &  -0.020 \\
0.006 & 	0.7684 & &0.8746 & & 3.4897 & & 3.9811 & & -0.143 & &  -0.082 \\
0.008& 0.8263& &0.9758 & & 	4.1786 & & 4.5970	& & -0.066 & & -0.044\\ 
0.01 & 0.8610 & &1.0374 & & 	4.7691& & 5.1396 & & -0.037 & & -0.027\\ \hline
0.0025 & 0.5798	& & 0.4682 & & 2.4901 & & 3.2702 & & -0.462 & & -0.150\\
0.0035 & 0.6999 & &0.7722 & &  3.2370 & & 3.8693 & & -0.157 & &  -0.070\\
0.0055 & 0.8090& & 0.9369 & & 4.3628	 & & 4.8504 & & -0.040 & &  -0.023\\
0.0075 &  0.8599 & &1.0018 & &  5.2525 & & 5.6641 & & -0.016 & &  -0.010\\
0.0085 &  0.8764	 & &1.0103 & & 4.2338& & 4.5224 & & -0.076 & & -0.058\\
\hline\hline
\end{tabular}
\caption{\label{fig:FVcorr} Results for the ratios $r_{\rm logs}$, $r(M_0L)$ and $r_{M_{\pm}}L$. Based on the parameters obtained in the combined fit V.}
\end{table}

In section \ref{ssect:FV} we argued that the FV corrections may be significantly larger for neutral pion masses much smaller than the charged ones. Having performed fits to the data we can {\em a posteriori} quantify this enhancement. We define the ratio
\begin{align}
r_{\rm logs} &= \frac{M_0^2\Big(\ln(M_0^2/\Lambda_3^2) - \tilde{g}_1(M_0L)\Big)}{M_{\pm}^2\Big(\ln(M_{\pm}^2/\Lambda_3^2)- \tilde{g}_1(M_{\pm}L)\Big)}
\end{align} 
of the neutral pion chiral log and the charged pion chiral log (with FV corrections included), and the ratio
\begin{align}
r(ML) &= \frac{\tilde{g}_1(ML)}{\ln (M^2/\Lambda_3^2)}
\end{align}
as a measure about the relative size of the FV correction to the (infinite volume) chiral log.

Table \ref{fig:FVcorr} gives the results for these ratios for the fit parameter $B_0,\Lambda_3$ and $c_2$ of the combined fit V. Apparently, for the smallest quark mass ($a\mu_0=0.0025$) the neutral pion chiral log is less than half as large as the charged pion chiral log. And $r(M_0L)$ is three times larger than $r(M_{\pm}L)$. Although less severe we find a significant increase in the FV corrections for the heavier quark masses as well.

The ETM collaboration has found in ref.\ \cite{Baron:2010bv} that the resummed FV corrections of \cite{Colangelo:2005gd} describe the data better than the standard 1-loop FV corrections with the charged pion mass. This might not be an indication for the superiority of the resummed formulae. It may just signal the failure of the standard expressions due to the use of the heavier charged pion mass.  

\section{Concluding remarks}

As already mentioned, the numerical analysis of the previous section can certainly be refined in various ways. For example, the error analysis should be improved, and the proper ratio of $Z$-factors should be included instead of the approximate value 1 employed here. Nevertheless, the results of the previous section show strong evidence for the sensitivity of the data on the pion mass splitting in the chiral logs. The fits are in general better with the mass splitting  taken into account.  Moreover, provided the general trends of table \ref{fig:combFit} persist, there is some sizable and non-vanishing impact of our new formulae on the extraction of LECs, in particular for $f$ and $\lt$.
We also found that separate fits at fixed lattice spacings can be quite misleading concerning the scaling violations in these LECs. These findings, if corroborated, are of course very important for ChPT phenomenology.

Further improvement seems possible if data for the neutral pion mass are included in the simultaneous fit. Although the neutral pion mass data are in general afflicted with significantly larger errors, including the data may still reduce the error on $c_2$ together with improved estimates for the other fit parameters, in particular the physical LECs.

An immediate question is whether more observables are affected in an analogous way by a  large pion mass splitting. Other purely pionic observables are pion scattering lengths. These have been studied in twisted mass WChPT \cite{Buchoff:2008hh}, but only in the GSM regime where the pion mass splitting is a NLO effect and therefore ignored in the pion loops.

More relevant in practice is the extension of the results given here to WChPT including the kaons. After all, the simulations in ref.\ \cite{Baron:2010bv} take into account a dynamical strange quark and we expect similar modifications in the 1-loop results for the mass and decay constant of the kaon. 

\section*{Acknowledgments}

Correspondence with  C.\ Urbach on the results in ref.\ \cite{Baron:2010bv} is gratefully acknowledged. I thank M.\ Golterman and K.\ Jansen for their feedback on the first draft of this manuscript.

This work is supported in part by the Deutsche Forschungsgemeinschaft (SFB/TR 09).

\end{document}